\pdfoutput=1 

\documentclass[reprint,aps,prc,showpacs,longbibliography]{revtex4-1}

\usepackage{amsmath,amssymb}
\usepackage[colorlinks=true,linkcolor=blue,urlcolor=blue,citecolor=blue]{hyperref}
\usepackage{graphicx}
\usepackage{multirow}

\DeclareMathOperator{\dilog}{\Phi}

\begin{document}

\title{Reanalysis of Rosenbluth measurements of the proton form factors}

\author{A.~V.~Gramolin}\email[Corresponding author: ]{gramolin@inp.nsk.su}
\author{D.~M.~Nikolenko}

\affiliation{Budker Institute of Nuclear Physics, 630090 Novosibirsk, Russia}

\begin{abstract}
We present a reanalysis of the data from Stanford Linear Accelerator Center (SLAC) experiments E140 [R.\,C.~Walker \textit{et~al.}, \href{http://dx.doi.org/10.1103/PhysRevD.49.5671}{Phys. Rev. D \textbf{49}, 5671 (1994)}] and NE11 [L.~Andivahis \textit{et~al.}, \href{http://dx.doi.org/10.1103/PhysRevD.50.5491}{Phys. Rev. D \textbf{50}, 5491 (1994)}] on elastic electron-proton scattering. This work is motivated by recent progress in calculating the corresponding radiative corrections and by the apparent discrepancy between the Rosenbluth and polarization transfer measurements of the proton electromagnetic form factors. New, corrected values for the scattering cross sections are presented, as well as a new form factor fit in the $Q^2$~range from 1 to~$8.83~\text{GeV}^2$. We also provide a complete set of revised formulas to account for radiative corrections in single-arm measurements of unpolarized elastic electron-proton scattering.
\end{abstract}


\maketitle

\section{Introduction}

The proton is an essential constituent of all atomic nuclei. Its static properties, including mass, electric charge, and magnetic moment, have been measured precisely~\cite{PDG2014}. In contrast, the proton's electromagnetic form factors, its fundamental dynamic characteristics, are still the subject of much ongoing research~\cite{ARNPS.54.217, JPhysG.34.S23, PPNP.59.694, PhysRep.550.1, EPJA.51.79}.

The proton electric and magnetic spacelike form factors, $G_E (Q^2)$ and $G_M (Q^2)$, are real-valued functions of the four-momentum transfer squared, $Q^2 = -q^2 \geqslant 0$, related to the spatial distributions of the electric charge and magnetic moment inside the proton. These can be measured in elastic lepton-proton scattering experiments. Starting from the pioneering work of Hofstadter~\cite{RMP.28.214} and up to the 1990s, the only method available was the Rosenbluth separation technique utilizing unpolarized $e^{-}p$~scattering. An alternative approach, the so-called polarization transfer method, is to measure the ratio $G_E / G_M$ using polarization observables. This was proposed back in 1968~\cite{Akhiezer&Rekalo(1968), Akhiezer&Rekalo(1974)}, but became available only recently with the development of intense polarized electron beams and recoil proton polarimeters. Surprisingly, the polarization transfer measurements~\cite{PRL.84.1398, PRC.71.055202, PRL.88.092301, PRC.85.045203, PRL.104.242301} yielded results contradicting the well-established data obtained with the Rosenbluth method. It was found that the discrepancy between the two sets of data rises with increasing four-momentum transfer.

As the Rosenbluth technique is much more sensitive to radiative corrections (RCs) than the polarization transfer method, this apparent contradiction could be explained by the neglected hard two-photon exchange (TPE) contribution to the elastic $e^{-}p$~scattering cross section~\cite{PPNP.66.782}. Although the recent experimental data~\cite{PRL.114.062005, PRL.114.062003} support this explanation, they were obtained for $Q^2 < 2~\text{GeV}^2$ while the discrepancy is significant only at higher four-momentum transfers. It has been alternatively proposed that inaccurate bremsstrahlung corrections are responsible for the discrepancy rather than the hard TPE effect. For example, the authors of Refs.~\cite{PRC.75.015207, PPNL.4.281} claim that the conflicting form factor measurements can be brought into agreement if the so-called structure function method is used to account for real photon emission. Because the proton form factor puzzle remains unsolved, it is important to consider all possibilities and to reexamine RCs applied in past Rosenbluth extractions of the proton form factors.

Unfortunately, for most of the measurements there is not sufficient information to perform an independent analysis of their RCs. Two notable exceptions are the Stanford Linear Accelerator Center (SLAC) experiments E140~\cite{PRD.49.5671, Walker_thesis} and NE11~\cite{PRD.50.5491, Clogher_thesis}, covering together the $Q^2$ range from~$1$ to~$8.83~\text{GeV}^2$. Both groups applied the same RC procedure, based on the standard prescription~\cite{RMP.41.205}, but including additional improvements presented in Ref.~\cite{PRD.49.5671}. Note that this procedure was later used again in the experiment~\cite{PRC.70.015206} performed at Jefferson Lab. However, both the original~\cite{RMP.41.205} and additional~\cite{PRD.49.5671} RC formulas contain misprints and inaccuracies, whose effects on the measurement results have never been investigated. To fill this gap, we revisit the RCs applied in Refs.~\cite{PRD.49.5671, PRD.50.5491} and perform a new extraction of the proton electric and magnetic form factors.

The paper is organized as follows. In Sec.~\ref{Sec2}, the basic formulas describing the unpolarized elastic $e^{-}p$~scattering are recalled. Section~\ref{Sec3} reviews the corresponding RCs and may be of independent interest. In Sec.~\ref{Sec4} we reanalyze the SLAC measurements and present our results. Finally, Sec.~\ref{Sec5} summarizes this work and the conclusions drawn from its results.

Throughout the paper we use a natural system of units where $\hbar = c = 1$ and the fine-structure constant is $\alpha = e^2 / (4\pi) \approx 1 / 137$. With this choice of units, all energies, momenta, and masses are expressed in GeV and scattering cross sections in $\text{GeV}^{-2}$ ($1~\text{GeV}^{-2} \approx 0.389~\text{mb}$). All formulas are written in the laboratory frame and neglecting the mass of the electron compared to its energy.

\vfill

\section{Unpolarized elastic electron-proton scattering}
\label{Sec2}

The differential cross section for unpolarized elastic electron-proton scattering is given in the lowest order in~$\alpha$ by the Rosenbluth formula
\begin{equation}
\frac{d \sigma_0}{d \Omega} = \frac{1}{\varepsilon (1 + \tau)} \left[\varepsilon G_E^2 (Q^2) + \tau G_M^2 (Q^2)\right] \frac{d \sigma_{\text{Mott}}}{d \Omega}, \label{Eq1}
\end{equation}
where
\begin{gather}
Q^2 = \frac{2M E_1^2 (1 - \cos{\theta})}{M + E_1 (1 - \cos{\theta})}, \quad \quad \tau = \frac{Q^2}{4M^2}, \\
\varepsilon = \left[1 + 2(1 + \tau) \tan^2{\frac{\theta}{2}}\right]^{-1}
\end{gather}
is the virtual-photon polarization parameter, $M$~is the proton mass, $E_1$~is the beam energy, and $\theta$~is the electron scattering angle. The Mott differential cross section, $d \sigma_{\text{Mott}} / d \Omega$, describes the scattering of electrons on spinless point particles of charge~$Z$ and is given by
\begin{equation}
\frac{d \sigma_{\text{Mott}}}{d \Omega} = \frac{Z^2 \alpha^2}{4 E_1^2} \frac{\cos^2{(\theta / 2)}}{\sin^4{(\theta / 2)}} \eta^{-1},
\end{equation}
where
\begin{equation}
\eta = 1 + \frac{E_1}{M} (1 - \cos{\theta})
\end{equation}
is the recoil factor. Though $Z = 1$ in the case of $e^{-}p$~scattering, we keep it for completeness.

The combination $\varepsilon G_E^2 + \tau G_M^2$ appearing in Eq.~(\ref{Eq1}) is often called the reduced cross section. Its linear dependence on~$\varepsilon$ forms the basis for the Rosenbluth separation technique. By varying beam energies and scattering angles, one can measure the reduced cross section at a fixed~$Q^2$, but for different values of~$\varepsilon$. Then, performing a linear fit of these cross-section data as a function of~$\varepsilon$, one determines $G_E^2$ as the slope and $\tau G_M^2$ as the intercept.

Based on numerous Rosenbluth measurements, it was established that the proton form factors approximately follow the dipole parametrization
\begin{equation}
G_E (Q^2) \approx G_D (Q^2), \quad G_M (Q^2) \approx \mu G_D (Q^2),
\end{equation}
where
\begin{equation}
G_D (Q^2) = \left(1 + \frac{Q^2}{\Lambda^2}\right)^{-2} \label{Eq7}
\end{equation}
is the dipole form factor, $\Lambda^2 = 0.71~\text{GeV}^2$, and $\mu \approx 2.79$~is the proton magnetic moment in units of the nuclear magneton.

\section{Radiative corrections to electron-proton scattering}
\label{Sec3}

A measured elastic scattering cross section inevitably contains the contributions of higher-order QED processes and therefore differs from that of Eq.~(\ref{Eq1}). The measured and Rosenbluth cross sections can be related by
\begin{equation}
\frac{d \sigma_{\text{meas}}}{d \Omega} = C_{\text{rad}} \frac{d \sigma_0}{d \Omega}, \label{Eq8}
\end{equation}
where $C_{\text{rad}}$ is the RC factor.

The RCs represented in Eq.~(\ref{Eq8}) by $C_{\text{rad}}$ are divided into two categories: internal and external. The former arise from the exchange of additional virtual photons and the emission of real photons during the act of electron-proton scattering~\cite{JPhysG.41.115001}. External RCs are due to bremsstrahlung and ionization processes accompanying the passage of the incident and outgoing particles through the target materials.

In general, RCs depend on the scattering kinematics, specific experimental conditions, and exact event selection. For this reason, accounting for RCs in coincidence experiments usually requires performing realistic Monte Carlo simulations~\cite{JPhysG.41.115001}. However, in a single-arm experiment where only electrons scattered at a fixed angle~$\theta$ are detected, the event selection procedure can be characterized by a single cut value, $\Delta E$, requiring that
\begin{equation}
E_3^{\text{el}} - E_3 \leqslant \Delta E, \label{Eq9}
\end{equation}
where $E_3^{\text{el}} = E_1 / \eta$ is the elastic peak energy and $E_3$~is the measured energy of the scattered electron. The latter is smaller than $E_3^{\text{el}}$ because of inelastic processes accompanying the elastic scattering. In this paper, we consider only the case of a single-arm experiment (performed, typically, with a high-resolution magnetic spectrometer).

To leading order in~$\alpha$, the RC factor is
\begin{equation}
C_{\text{rad}} = 1 + \delta (\Delta E), \label{Eq10}
\end{equation}
where $\delta (\Delta E)$ can be calculated using various theoretical prescriptions. The most commonly applied is that given by Mo and Tsai in 1969~\cite{RMP.41.205}. More recently, Maximon and Tjon~\cite{PRC.62.054320} provided another prescription, removing several mathematical and physical approximations used by Mo and Tsai. We refer to the calculations~\cite{RMP.41.205, PRC.62.054320} as the standard RC prescriptions.

The factor~(\ref{Eq10}) accounts only for the lowest-order RCs, when bremsstrahlung is reduced to the emission of a single photon. As shown by Yennie, Frautschi, and Suura~\cite{AnnPhys.13.379}, the emission of an arbitrary number of soft photons can be taken into account via the exponentiation of~$\delta (\Delta E)$:
\begin{equation}
C_{\text{rad}} = \exp{\left[\delta (\Delta E)\right]}. \label{Eq11}
\end{equation}
The difference between the values of~$C_{\text{rad}}$ given by Eqs. (\ref{Eq10}) and~(\ref{Eq11}) increases with decreasing $\Delta E$ and can be large for experiments using high-resolution detectors.

In this paper we adopt the definition of~$C_{\text{rad}}$ similar to that used in Refs.~\cite{PRD.49.5671, PRD.50.5491}:
\begin{equation}
C_{\text{rad}} = \exp{(\delta_{\text{MTj}} + \delta_{\text{vac}} + \delta_{\text{int.br.}} + \delta_{\text{ext.br.}})} C_L, \label{Eq12}
\end{equation}
where $\delta_{\text{MTj}}$ represents the standard RCs according to Maximon and Tjon~\cite{PRC.62.054320}, $\delta_{\text{vac}}$ is the part of the vacuum polarization correction unaccounted for by the standard prescriptions, $\delta_{\text{int.br.}}$ is an additional correction helping to improve the description of hard internal bremsstrahlung, $\delta_{\text{ext.br.}}$ is the term accounting for external bremsstrahlung, and $C_L$ is the correction factor due to ionization losses in the target materials. It has been argued that the vacuum polarization correction is infrared finite and therefore should not be exponentiated. However, this does not lead to a significant change in the numerical value of~$C_{\text{rad}}$ since $\delta_{\text{vac}}$ is small compared to the other terms exponentiated.

The terms $\delta_{\text{MTj}}$, $\delta_{\text{vac}}$, and $\delta_{\text{int.br.}}$ correspond to internal RCs, while $\delta_{\text{ext.br.}}$ and~$C_L$ represent the external ones. We discuss each of these contributions separately in the following subsections.

\begin{widetext}
\subsection{Standard radiative corrections}

In the measurements~\cite{PRD.49.5671, PRD.50.5491} discussed, Eq.~(II.6) of Mo and Tsai~\cite{RMP.41.205} was used to account for the standard RCs. In contrast, we use the following correction derived more recently by Maximon and Tjon~\cite{PRC.62.054320}:
\begin{gather}
\delta_{\text{MTj}} = \frac{\alpha}{\pi} \left[\frac{13}{6} \ln{\frac{Q^2}{m^2}} - \frac{28}{9} - \Bigl(\ln{\frac{Q^2}{m^2}} - 1\Bigr) \ln{\frac{4E_1 E_3^{\text{el}}}{(2\eta \Delta E)^2}} - \frac{1}{2} \ln^2{\eta} + \dilog{\Bigl(\cos^2{\frac{\theta}{2}}\Bigr)} - \frac{\pi^2}{6}\right] \nonumber \\
{} + \frac{2\alpha Z}{\pi} \left[-\ln{\eta} \, \ln{\frac{Q^2 x}{(2\eta \Delta E)^2}} + \dilog{\Bigl(1 - \frac{\eta}{x}\Bigr)} - \dilog{\Bigl(1 - \frac{1}{\eta x}\Bigr)}\right] + \frac{\alpha Z^2}{\pi} \left\{\frac{E_4}{|\mathbf{p}_4|} \left[-\frac{1}{2} \ln^2{x} - \ln{x} \, \ln{\frac{Q^2 + 4M^2}{M^2}} \right. \right. \nonumber \\
\left. {} + \ln{x} -\dilog{\Bigl(1 - \frac{1}{x^2}\Bigr)} + 2\dilog{\Bigl(-\frac{1}{x}\Bigr)} + \frac{\pi^2}{6}\right] - \left. \Bigl(\frac{E_4}{|\mathbf{p}_4|} \ln{x} - 1\Bigr) \ln{\frac{M^2}{(2\eta \Delta E)^2}} + 1\right\}, \label{Eq13}
\end{gather}
\end{widetext}
where $m$ is the electron mass, $E_4 = M + E_1 - E_3^{\text{el}}$ and $|\mathbf{p}_4| = \sqrt{E_4^2 - M^2}$ are the energy and momentum of the recoil proton, and $x = (E_4 + |\mathbf{p}_4|) / M$. The function $\dilog$ is Spence's function (or dilogarithm), defined as
\begin{equation}
\dilog{(y)} = -\int\limits_{0}^{y} \frac{\ln{|1 - u|}}{u} \, du.
\end{equation}

The Mo--Tsai~\cite{RMP.41.205} and Maximon--Tjon~\cite{PRC.62.054320} calculations differ in three major aspects (see Ref.~\cite{PAN.78.69} for a detailed discussion). First, an additional term, $\delta_{\text{el}}^{(1)}$, was introduced in Ref.~\cite{PRC.62.054320} to better account for the proton vertex correction. We have neglected this term in Eq.~(\ref{Eq13}) because it is small and model dependent on the proton form factors. Second, two different parametrizations of the soft TPE terms were used in Refs.~\cite{RMP.41.205} and~\cite{PRC.62.054320}. It is a matter of convention which definition to use. To switch from the Maximon--Tjon prescription for the soft TPE terms to the Mo--Tsai prescription, one should subtract from Eq.~(\ref{Eq13}) the following correction~\cite{JPhysG.41.115001}:
\begin{gather}
\delta_{2\gamma}' = -\frac{\alpha Z}{\pi} \left[\ln{\eta} \ln{\frac{Q^4}{4M^2 E_1 E_3^{\text{el}}}} + 2\dilog{\left(1 - \frac{M}{2E_1}\right)} \right. \nonumber \\
\left. {} - 2\dilog{\left(1 - \frac{M}{2E_3^{\text{el}}}\right)}\right].
\end{gather}
Finally, both groups of authors rely on the same assumptions while calculating the soft bremsstrahlung terms, but their results are different. The reason for this was identified in Ref.~\cite{PAN.78.69} as an incorrect substitution made by Mo and Tsai~\cite{RMP.41.205}.

\subsection{Vacuum polarization}

The contributions from virtual $e^+ e^-$, $\mu^+ \mu^-$, and $\tau^+ \tau^-$ loops to the vacuum polarization correction are described by the general formula
\begin{gather}
\delta_{\text{vac}}^{e, \mu, \tau} = \frac{2\alpha}{3\pi} \left\{-\frac{5}{3} + \frac{4m_{\ell}^2}{Q^2} + \left(1 - \frac{2m_{\ell}^2}{Q^2}\right) \sqrt{1 + \frac{4 m_{\ell}^2}{Q^2}} \rule{0mm}{8mm}\right. \nonumber \\
\left. \times \ln{\left[\frac{Q^2}{4 m_{\ell}^2} \left(1 + \sqrt{1 + \frac{4 m_{\ell}^2}{Q^2}}\right)^{\!\!\!2}\right]}\right\}, \label{Eq16}
\end{gather}
where $m_{\ell}$~is the mass of the corresponding lepton (electron, muon, or tau). Note that Eq.~(\ref{Eq16}) is different from the misprinted Eq.~(A5) in Ref.~\cite{PRD.49.5671}. Typically, $Q^2 \gg m^2$ (recall that $m$~is the electron mass) and the correction due to $e^+ e^-$ loops can be simplified to
\begin{equation}
\delta_{\text{vac}}^{e} = \frac{2\alpha}{3\pi} \left(-\frac{5}{3} + \ln{\frac{Q^2}{m^2}}\right). \label{Eq17}
\end{equation}
The contribution~(\ref{Eq17}) is already taken into account in the standard RC prescriptions and, in particular, it is included in Eq.~(\ref{Eq13}).

The hadronic part of the vacuum polarization correction, $\delta_{\text{vac}}^q$, cannot be calculated from first principles, but can be reliably extracted from experimental data on the annihilation of $e^+ e^-$ into hadrons. We use the same parametrization as that given by Eq.~(A6) in Ref.~\cite{PRD.49.5671}:
\begin{equation}
\delta_{\text{vac}}^{q} = 0.002 \left[1.513 + 2.822 \ln{\bigl(1 + 1.218 Q^2\bigr)}\right].
\end{equation}

Summing up, the total vacuum polarization correction unaccounted for in Eq.~(\ref{Eq13}) is
\begin{equation}
\delta_{\text{vac}} = \delta_{\text{vac}}^{\mu} + \delta_{\text{vac}}^{\tau} + \delta_{\text{vac}}^{q}.
\end{equation}

\subsection{More accurate description of hard internal bremsstrahlung}

Differentiating Eq.~(\ref{Eq13}) with respect to~$\Delta E$, we obtain
\begin{gather}
\frac{\partial \delta_{\text{MTj}}}{\partial (\Delta E)} = \frac{2\alpha}{\pi} \frac{1}{\Delta E} \left[\ln{\frac{Q^2}{m^2}} - 1 + 2Z \ln{\eta} \right. \nonumber \\
\left. {} + Z^2 \left(\frac{E_4}{|\mathbf{p}_4|} \ln{x} - 1\right)\right].
\end{gather}
Then, taking into account Eq.~(\ref{Eq9}), we can write the following differential cross section describing the radiative tail due to internal bremsstrahlung:
\begin{gather}
\frac{d^2 \sigma_{\text{int.br.}}}{d \Omega \, d E_3} = \frac{2\alpha}{\pi} \frac{1}{E_3^{\text{el}} - E_3} \left[\ln{\frac{Q^2}{m^2}} - 1 + 2Z \ln{\eta} \right. \nonumber \\
\left. {} + Z^2 \left(\frac{E_4}{|\mathbf{p}_4|} \ln{x} - 1\right)\right] \frac{d \sigma_0}{d \Omega}. \label{Eq21}
\end{gather}
The terms proportional to $Z^0$, $Z^2$, and~$Z^1$ represent, respectively, the electron bremsstrahlung, the proton bremsstrahlung, and the interference between them. Note that an equivalent expression follows also from Eq.~(II.6) of Mo and Tsai~\cite{RMP.41.205}.

\begin{figure}
\includegraphics[width=\columnwidth]{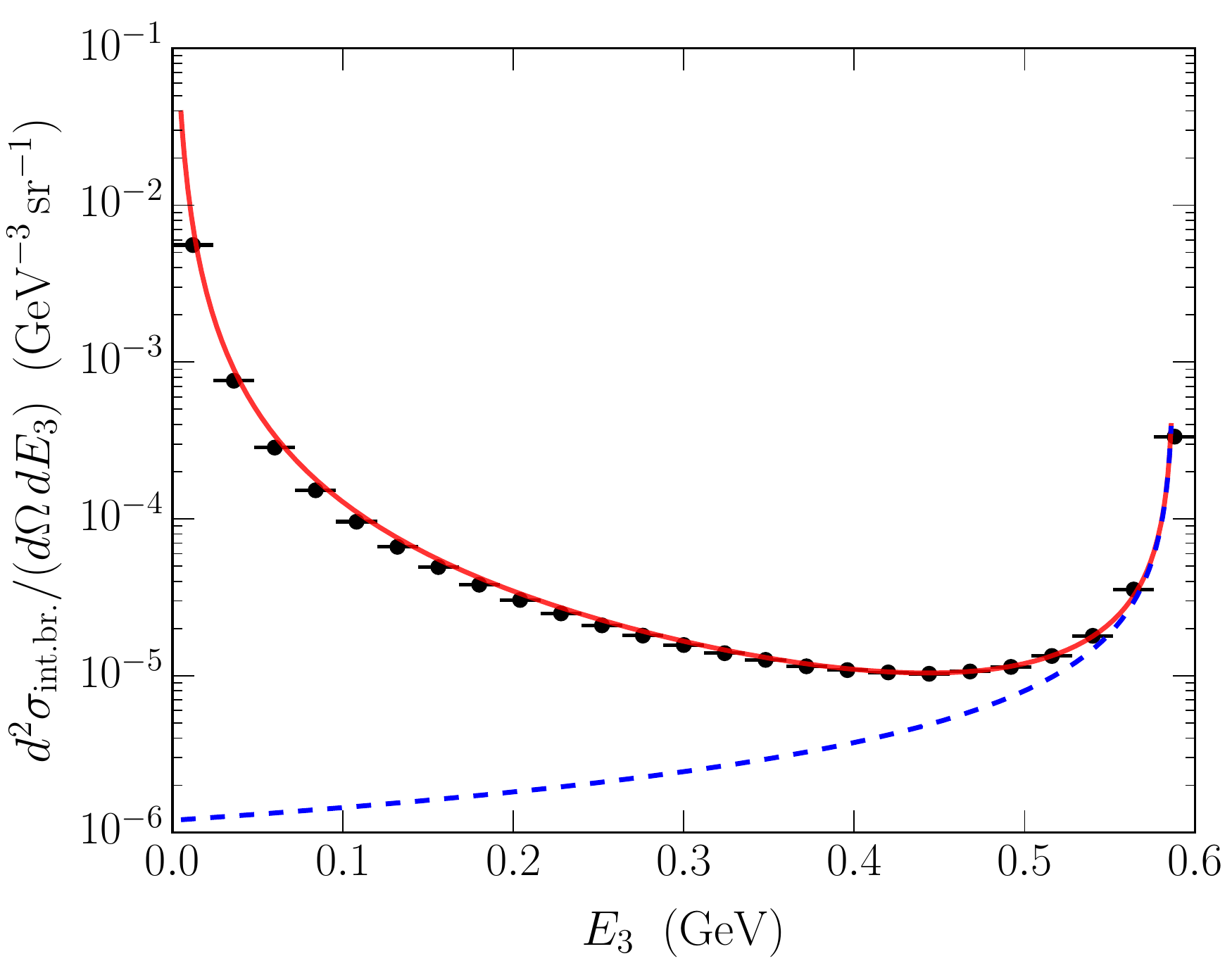}
\caption{\label{Fig1}Three different predictions for the radiative tail in the case when $E_1 = 1~\text{GeV}$, $\theta = 70^{\circ}$, and assuming dipole form factors. The elastic peak energy is $E_3^{\text{el}} = 0.588~\text{GeV}$. The blue dashed line represents the cross section~(\ref{Eq21}) based on the soft-photon approximation. The red solid line provides a better description of the radiative tail and is given by the sum of Eq.~(\ref{Eq22}) with the terms from Eq.~(\ref{Eq21}) proportional to~$Z$ and~$Z^2$. The points are simulated using the ESEPP (elastic scattering of electrons and positrons on protons) event generator~\cite{JPhysG.41.115001} with the accurate QED model.}
\end{figure}

Both Eq.~(\ref{Eq13}) and its counterpart (\ref{Eq21}) are valid only in the soft-photon approximation, i.e., assuming that the emission of a bremsstrahlung photon does not affect the elastic cross section $d \sigma_0 / d \Omega$. However, if the incident electron emits a hard photon and thus loses a sufficient part of its energy, the probability of a subsequent scattering on the proton increases~\cite{RMP.41.205, JPhysG.41.115001}. This can lead to the substantial growth of the cross section with increasing energy of the bremsstrahlung photon or, in other words, to a large rise in the radiative tail at low energies (see Fig.~\ref{Fig1}).

To account for the kinematic effect discussed, we use Eq.~(C.11) proposed by Mo and Tsai~\cite{RMP.41.205}, which describes electron bremsstrahlung in the peaking approximation:
\begin{gather}
\frac{d^2 \sigma_{\text{int.br.}}}{d \Omega \, d E_3} = \frac{M + (E_1 - \omega_1) (1 - \cos{\theta})}{M - E_3 (1 - \cos{\theta})} \nonumber \\
{} \times \frac{t_1}{\omega_1} \frac{d \sigma_0}{d \Omega} (E_1 - \omega_1) + \frac{t_3}{\omega_3} \frac{d \sigma_0}{d \Omega} (E_1), \label{Eq22}
\end{gather}
where
\begin{gather}
t_{1,3} = \frac{\alpha}{\pi} \left[\frac{1 + x_{1,3}^2}{2} \ln{\frac{2E_1 E_3 (1 - \cos{\theta})}{m^2}} - x_{1,3}\right], \\
x_1 = \frac{E_1 - \omega_1}{E_1}, \quad \quad x_3 = \frac{E_3}{E_3 + \omega_3}, \\
\omega_1 = R \, \omega_3, \quad \quad \omega_3 = E_3^{\text{el}} - E_3, \\
R = \frac{M + E_1 (1 - \cos{\theta})}{M - E_3 (1 - \cos{\theta})}.
\end{gather}
Here, $\omega_1$ ($\omega_3$) is the energy of the bremsstrahlung photon emitted in the direction of the incident (scattered) electron and $R$~is the ratio of~$\omega_1$ to~$\omega_3$. The first and the second terms in Eq.~(\ref{Eq22}) are due to bremsstrahlung by the incident and the scattered electrons, respectively. Note that in the soft-photon limit, when $\omega_{1,3} \ll E_{1,3}$, the differential cross section~(\ref{Eq22}) reduces to
\begin{equation}
\frac{d^2 \sigma_{\text{int.br.}}}{d \Omega \, d E_3} = \frac{2\alpha}{\pi} \frac{1}{E_3^{\text{el}} - E_3} \left(\ln{\frac{Q^2}{m^2}} - 1\right) \frac{d \sigma_0}{d \Omega} (E_1), \label{Eq27}
\end{equation}
which coincides with the terms in Eq.~(\ref{Eq21}) describing electron bremsstrahlung.

The resulting additional correction to Eq.~(\ref{Eq13}) can be written as
\begin{gather}
\delta_{\text{int.br.}} = \frac{2\alpha}{\pi} \left(\ln{\frac{Q^2}{m^2}} - 1\right) \ln{\frac{\delta E}{\Delta E}} \nonumber \\
{} + \left[\frac{d \sigma_0}{d \Omega} (E_1)\right]^{-1} \int\limits_{E_3^{\text{el}} - \Delta E}^{E_3^{\text{el}} - \delta E} \frac{d^2 \sigma_{\text{int.br.}}}{d \Omega \, d E_3} \, d E_3, \label{Eq28}
\end{gather}
where the integrand is given by Eq.~(\ref{Eq22}) and the integration can be performed numerically. As expected, the value $\delta_{\text{int.br.}} = 0$ is obtained when using the cross section~(\ref{Eq27}) as the integrand. The cutoff energy $\delta E$ should be chosen so that $\delta E \ll \Delta E$. In our analysis, we use the value $\delta E = 10^{-4}~\text{GeV}$.

Figure~\ref{Fig1} compares two analytical descriptions and a numerical calculation of the radiative tail for certain kinematics. It can be seen that the cross section given by Eq.~(\ref{Eq21}) decreases monotonically with decreasing~$E_3$. A more accurate analytical description is obtained by combining Eq.~(\ref{Eq22}) with the terms from Eq.~(\ref{Eq21}) proportional to~$Z$ and~$Z^2$. This corresponds to the correction~(\ref{Eq28}) and is in good agreement with the data points simulated using the \mbox{ESEPP} event generator~\cite{JPhysG.41.115001}.

\subsection{External bremsstrahlung}

To calculate the radiative tail due to external bremsstrahlung, we use the following cross section similar to that given by Eq.~(C.13) in Ref.~\cite{SLAC-PUB-848}:
\begin{gather}
\frac{d^2 \sigma_{\text{ext.br.}}}{d \Omega \, d E_3} = \frac{1}{\Gamma (1 + b_i t_i)} \frac{1}{\Gamma (1 + b_f t_f)} \Bigl(\frac{\omega_1}{E_1}\Bigr)^{b_i t_i} \Bigl(\frac{\omega_3}{E_3^{\text{el}}}\Bigr)^{b_f t_f} \nonumber \\
{} \times \left[\frac{M + (E_1 - \omega_1) (1 - \cos{\theta})}{M - E_3 (1 - \cos{\theta})} \frac{b_i t_i}{\omega_1} \phi \Bigl(\frac{\omega_1}{E_1}\Bigr) \right. \nonumber \\
\left. {} \times \frac{d \sigma_0}{d \Omega} (E_1 - \omega_1) + \frac{b_f t_f}{\omega_3} \phi \Bigl(\frac{\omega_3}{E_3^{\text{el}}}\Bigr) \frac{d \sigma_0}{d \Omega} (E_1)\right], \label{Eq29}
\end{gather}
where the function
\begin{equation}
\phi \Bigl(\frac{\omega}{E}\Bigr) = 1 - \frac{\omega}{E} + \frac{3}{4} \Bigl(\frac{\omega}{E}\Bigr)^2
\end{equation}
describes the shape of the bremsstrahlung spectrum in the complete screening case and normalized such that $\phi (0) = 1$. As usual, $\Gamma$ denotes the gamma function. The quantity $t_{i,f}$ represents the thickness of the material expressed in units of its radiation length, $X_0$, where the subscripts~$i$ and~$f$ refer to the materials traversed by the incident and scattered electrons, respectively. The dimensionless parameter~$b_{i,f}$ is~\cite{RMP.46.815}
\begin{equation}
b = \frac{4}{3} + \frac{4}{9} \alpha r_e^2 N_A \frac{Z (Z + 1)}{A} X_0,
\end{equation}
where $r_e$~is the classical electron radius, $N_A$~is the Avogadro constant, $Z$~is the atomic number of the material, and $A$~is its atomic mass. Note that the factor $\alpha r_e^2 N_A$ is about $3.49 \times 10^{-4}~\text{cm}^2 / \text{mol}$. For hydrogen, since $Z = 1$, $A = 1.00794~\text{g} / \text{mol}$, and $X_0 = 63.04~\text{g} / \text{cm}^2$~\cite{PDG2014}, $b \approx 1.353$. For simplicity, in our analysis we use this value for both $b_i$ and~$b_f$.

The cross section~(\ref{Eq29}) is similar to that derived in Ref.~\cite{PRD.49.5671}. In fact, their Eq.~(A14) can be obtained from Eq.~(\ref{Eq29}) after substituting $\omega_1 = R \Delta E$ and $\omega_3 = \Delta E$ and using the approximation
\begin{equation}
\frac{M + (E_1 - \omega_1) (1 - \cos{\theta})}{M - E_3 (1 - \cos{\theta})} \approx R \approx \eta^2.
\end{equation}

In the vicinity of the elastic peak, the cross section~(\ref{Eq29}) can be integrated as
\begin{gather}
\int\limits_{E_3^{\text{el}} - \delta E}^{E_3^{\text{el}}} \frac{d^2 \sigma_{\text{ext.br.}}}{d \Omega \, d E_3} \, d E_3 = \frac{1}{\Gamma (1 + b_i t_i)} \frac{1}{\Gamma (1 + b_f t_f)} \nonumber \\
\times \left(\frac{\eta^2 \delta E}{E_1}\right)^{\!b_i t_i} \left(\frac{\delta E}{E_3^{\text{el}}}\right)^{\!b_f t_f} \frac{d \sigma_0}{d \Omega} (E_1).
\end{gather}
This allows us to finally write the following expression for the RCs due to external bremsstrahlung:
\begin{gather}
\exp{(\delta_{\text{ext.br.}})} = \frac{1}{\Gamma (1 + b_i t_i)} \frac{1}{\Gamma (1 + b_f t_f)} \nonumber \\
\times \left(\frac{\eta^2 \delta E}{E_1}\right)^{\!b_i t_i} \left(\frac{\delta E}{E_3^{\text{el}}}\right)^{\!b_f t_f} \nonumber \\
{} + \left[\frac{d \sigma_0}{d \Omega} (E_1)\right]^{-1} \int\limits_{E_3^{\text{el}} - \Delta E}^{E_3^{\text{el}} - \delta E} \frac{d^2 \sigma_{\text{ext.br.}}}{d \Omega \, d E_3} \, d E_3. \label{Eq34}
\end{gather}
Note that the correction~(\ref{Eq34}) is already exponentiated.

\subsection{Ionization losses}

Another process that causes a decrease in energy of electrons passing through the target materials is the ionization and excitation of the target atoms. In the case of ultrarelativistic electrons, the most probable energy loss due to this process is given by the formula~\cite{PDG2014}
\begin{equation}
\Delta E_0 = \xi \left(\ln{\frac{\alpha^2 X_0 t}{r_e \rho}} + 0.2\right), \label{Eq35}
\end{equation}
where
\begin{equation}
\xi = 2\pi m r_e^2 N_A \frac{Z}{A} X_0 t, \label{Eq36}
\end{equation}
$X_0 t$ is the thickness of the material in~$\text{g} / \text{cm}^2$, and~$\rho$ is its density in~$\text{g} / \text{cm}^3$. The factor $2\pi m r_e^2 N_A$ in Eq.~(\ref{Eq36}) is about $1.535 \times 10^{-4}~\text{GeV} \, \text{cm}^2 / \text{mol}$.

The nominal beam energy and the measured energy of the scattered electron should be corrected for the corresponding values of the most probable energy loss. This has already been done in the discussed papers~\cite{PRD.49.5671, PRD.50.5491}, and the incident electron energies given there are after subtracting~$\Delta E_0$. Equation (11) from Ref.~\cite{PRD.49.5671} was used for this purpose, which is equivalent to Eq.~(\ref{Eq35}).

The energy loss due to ionization is subject to random fluctuations described by the Landau distribution~\cite{JPhysUSSR.8.201}:
\begin{equation}
L (\lambda) = \frac{1}{\pi} \int\limits_{0}^{\infty} \exp{(-u \ln{u} - \lambda u)} \sin{(\pi u)} \, du, \label{Eq37}
\end{equation}
where
\begin{equation}
\lambda = \frac{\Delta E_{\lambda} - \Delta E_0}{\xi}
\end{equation}
is a parameter characterizing the deviation of the actual energy loss~$\Delta E_{\lambda}$ from the most probable value~(\ref{Eq35}). For $\lambda \gg 1$, the distribution~(\ref{Eq37}) is $L (\lambda) \propto \lambda^{-2}$.

For the experiments under consideration, a correction due to Landau fluctuations is small and thus can be approximated as~\cite{PRD.49.5671}
\begin{equation}
C_L = 1 - \frac{1}{\xi_i} \int\limits_{\eta^2 \Delta E}^{\infty} L \Bigl(\frac{\omega}{\xi_i}\Bigr) d\omega - \frac{1}{\xi_f} \int\limits_{\Delta E}^{\infty} L \Bigl(\frac{\omega}{\xi_f}\Bigr) d\omega, \label{Eq39}
\end{equation}
where the subscripts~$i$ and~$f$ have the same meaning as above. Note that the corresponding Eq.~(A19) in Ref.~\cite{PRD.49.5671} contains misprints. The value of~$C_L$ represents the probability that the condition~(\ref{Eq9}) is satisfied despite the ionization losses. To calculate the integrals in Eq.~(\ref{Eq39}) more effectively, the following identity can be used~\cite{JPhysUSSR.8.201}:
\begin{equation}
\int\limits_{y}^{\infty} L(\lambda) \, d\lambda = \int\limits_{0}^{\infty} \exp{(-u \ln{u} - y u)} \frac{\sin{(\pi u)}}{\pi u} \, du.
\end{equation}

\section{Reanalysis of the SLAC measurements}
\label{Sec4}

Here we perform a reanalysis of the data collected in the SLAC experiments~\cite{PRD.49.5671} and~\cite{PRD.50.5491}. Walker~\textit{et~al.}~\cite{PRD.49.5671} measured the elastic $e^{-}p$ scattering cross sections for 22 different kinematics with the $Q^2$ values of $1$, $2.003$, $2.497$, and $3.007~\text{GeV}^2$. According to Ref.~\cite{PRC.68.034325}, the small-angle data of Ref.~\cite{PRD.49.5671} are not reliable because of a missing experimental correction and, therefore, should be excluded from the analysis. The remaining $N_1 = 16$ data points with $\theta > 15^{\circ}$ are referred to as ``Set~1'' (see Table~\ref{Tab1}).

Andivahis~\textit{et~al.}~\cite{PRD.50.5491} performed their measurements at the $Q^2$~values of $1.75$, $2.5$, $3.25$, $4$, $5$, $6$, $7$, and $8.83~\text{GeV}^2$. Two separate magnetic spectrometers were used in the experiment. The larger one detected electrons with momenta up to~$8~\text{GeV}$ and was rotated around the target pivot to set the scattering angle. The $N_2 = 24$ cross sections obtained with this apparatus are referred to as ``Set~2.'' The smaller spectrometer, operating at momenta up to $1.6~\text{GeV}$, was fixed at the angle $\theta = 90^{\circ}$. The corresponding $N_3 = 8$ data points comprise ``Set~3.''

The RCs discussed in Sec.~\ref{Sec3} are expressed through the cut value $\Delta E$ appearing in Eq.~(\ref{Eq9}). In contrast, a cut on the missing mass squared, $W^2 \leqslant W_{\text{cut}}^2$, was used to select elastic scattering events in the measurements reanalyzed. Let us show how the quantities $\Delta E$ and $W_{\text{cut}}^2$ are related.

The missing mass squared is, by definition, 
\begin{equation}
W^2 = (\ell_1 + p_2 - \ell_3)^2 = M^2 + 2\eta M (E_3^{\text{el}} - E_3),
\end{equation}
where $\ell_1$ ($\ell_3$) is the four-momentum of the incident (scattered) electron and $p_2$~is the four-momentum of the target proton. In particular, $W^2 = M^2$ in the case of purely elastic scattering, i.e., when $E_3 = E_3^{\text{el}}$. It can be easily seen that the condition~(\ref{Eq9}) is satisfied if
\begin{equation}
W_{\text{cut}}^2 = M^2 + 2\eta M \Delta E. \label{Eq42}
\end{equation}
In the experiments of interest, the values of $W_{\text{cut}}^2$ were in the range from 0.96 to $1.16~\text{GeV}^2$. The upper limit is approximately equal to the pion production threshold, $W^2 = (M + m_{\pi})^2$, where $m_{\pi}$ is the $\pi^0$~mass. The corresponding values of~$\Delta E$ can be calculated using Eq.~(\ref{Eq42}).

The numerical results of applying the RCs described in Sec.~\ref{Sec3} to the uncorrected data of the measurements~\cite{PRD.49.5671, PRD.50.5491} are shown in Table~\ref{Tab1}. The values of~$Q^2$ and~$\varepsilon$ listed there are nominal, while in the original analyses RCs were applied to the measured cross sections before converting them to the nominal kinematics. For this reason, we calculated the new RCs based on the actual values of~$E_1$ and~$\theta$ also provided in Refs.~\cite{PRD.49.5671, Walker_thesis, PRD.50.5491, Clogher_thesis}. All input data used and our {\scshape python} analysis routine are made freely available~\cite{GitHub}.

\begin{table*}
\begin{ruledtabular}
\caption{\label{Tab1}Radiative corrections and differential cross sections obtained by reanalysis of the measurements~\cite{PRD.49.5671, PRD.50.5491}.}
\begin{tabular}{cccccccccccccc}
\multirow{2}{*}{Set\!\!} & $Q^2$ & $\varepsilon$ & $\delta_{\text{MTj}}$ & $\delta_{\text{vac}}$ & $\delta_{\text{int.br.}}$ & $\delta_{\text{ext.br.}}$ & $C_L$ & $C_{\text{rad}}^{\text{new}}$ & \!\!\!$C_{\text{rad}}^{\text{old}} / C_{\text{rad}}^{\text{new}}$\!\!\! & $d \sigma_0 / d \Omega$ & $\Delta_{\text{stat}}$ & $\Delta_{\text{syst}}$ & $\Delta_{\text{norm}}$ \\
& ($\text{GeV}^2$) & & & & & & & & & ($\text{nb} / \text{sr}$) & (\%) & (\%) & (\%) \\
\hline
1 & 1.000 & 0.692 & $-0.1657$ & 0.0121 & 0.0050 & $-0.1098$ & 0.9934 & 0.7672 & 1.0045 & $5.291{\times}10^{+0}$ & 0.80 & 0.50 & 1.90 \\
1 & 1.000 & 0.869 & $-0.1690$ & 0.0122 & 0.0042 & $-0.1268$ & 0.9951 & 0.7525 & 1.0032 & $1.786{\times}10^{+1}$ & 0.91 & 0.50 & 1.90 \\
1 & 1.000 & 0.930 & $-0.1702$ & 0.0122 & 0.0039 & $-0.1410$ & 0.9960 & 0.7414 & 1.0023 & $3.960{\times}10^{+1}$ & 0.86 & 0.50 & 1.90 \\
1 & 2.003 & 0.635 & $-0.1802$ & 0.0157 & 0.0077 & $-0.1064$ & 0.9950 & 0.7648 & 1.0057 & $4.461{\times}10^{-1}$ & 0.92 & 0.50 & 1.90 \\
1 & 2.003 & 0.735 & $-0.1782$ & 0.0157 & 0.0073 & $-0.1125$ & 0.9958 & 0.7618 & 1.0038 & $7.827{\times}10^{-1}$ & 0.75 & 0.50 & 1.90 \\
1 & 2.003 & 0.808 & $-0.1791$ & 0.0158 & 0.0067 & $-0.1204$ & 0.9962 & 0.7552 & 1.0035 & $1.292{\times}10^{+0}$ & 0.61 & 0.50 & 1.90 \\
1 & 2.003 & 0.878 & $-0.1781$ & 0.0158 & 0.0064 & $-0.1286$ & 0.9969 & 0.7500 & 1.0025 & $2.427{\times}10^{+0}$ & 0.96 & 0.50 & 1.90 \\
1 & 2.003 & 0.938 & $-0.1966$ & 0.0158 & 0.0046 & $-0.1472$ & 0.9970 & 0.7216 & 1.0014 & $5.754{\times}10^{+0}$ & 2.38 & 0.50 & 1.90 \\
1 & 2.497 & 0.619 & $-0.1833$ & 0.0170 & 0.0089 & $-0.1043$ & 0.9956 & 0.7663 & 1.0054 & $1.904{\times}10^{-1}$ & 0.91 & 0.50 & 1.90 \\
1 & 2.497 & 0.723 & $-0.1760$ & 0.0170 & 0.0090 & $-0.1075$ & 0.9965 & 0.7703 & 1.0043 & $3.383{\times}10^{-1}$ & 0.85 & 0.50 & 1.90 \\
1 & 2.497 & 0.800 & $-0.1813$ & 0.0170 & 0.0078 & $-0.1146$ & 0.9968 & 0.7601 & 1.0033 & $5.648{\times}10^{-1}$ & 0.63 & 0.50 & 1.90 \\
1 & 2.497 & 0.846 & $-0.1818$ & 0.0171 & 0.0074 & $-0.1247$ & 0.9970 & 0.7520 & 1.0026 & $8.315{\times}10^{-1}$ & 0.93 & 0.50 & 1.90 \\
1 & 3.007 & 0.623 & $-0.1852$ & 0.0181 & 0.0100 & $-0.1043$ & 0.9961 & 0.7670 & 1.0056 & $9.719{\times}10^{-2}$ & 0.97 & 0.50 & 1.90 \\
1 & 3.007 & 0.761 & $-0.1829$ & 0.0182 & 0.0090 & $-0.1133$ & 0.9970 & 0.7618 & 1.0041 & $2.203{\times}10^{-1}$ & 0.85 & 0.50 & 1.90 \\
1 & 3.007 & 0.910 & $-0.2149$ & 0.0182 & 0.0049 & $-0.1569$ & 0.9968 & 0.7034 & 1.0019 & $9.102{\times}10^{-1}$ & 2.55 & 0.50 & 1.90 \\
1 & 3.007 & 0.932 & $-0.2250$ & 0.0183 & 0.0042 & $-0.1693$ & 0.9967 & 0.6872 & 1.0015 & $1.317{\times}10^{+0}$ & 1.05 & 0.50 & 1.90 \\
2 & 1.750 & 0.250 & $-0.1870$ & 0.0149 & 0.0087 & $-0.0653$ & 0.9928 & 0.7899 & 1.0090 & $1.453{\times}10^{-1}$ & 0.78 & 1.06 & 1.77 \\
2 & 1.750 & 0.704 & $-0.1786$ & 0.0149 & 0.0066 & $-0.0771$ & 0.9965 & 0.7884 & 1.0036 & $1.033{\times}10^{+0}$ & 0.46 & 1.06 & 1.77 \\
2 & 1.750 & 0.950 & $-0.1814$ & 0.0149 & 0.0049 & $-0.0941$ & 0.9983 & 0.7732 & 1.0014 & $1.157{\times}10^{+1}$ & 0.58 & 1.06 & 1.77 \\
2 & 2.500 & 0.227 & $-0.1877$ & 0.0170 & 0.0128 & $-0.0601$ & 0.9941 & 0.7994 & 1.0113 & $3.427{\times}10^{-2}$ & 1.07 & 1.06 & 1.77 \\
2 & 2.500 & 0.479 & $-0.1864$ & 0.0170 & 0.0097 & $-0.0674$ & 0.9960 & 0.7937 & 1.0066 & $9.922{\times}10^{-2}$ & 0.93 & 1.06 & 1.77 \\
2 & 2.500 & 0.630 & $-0.1866$ & 0.0170 & 0.0084 & $-0.0741$ & 0.9967 & 0.7877 & 1.0047 & $1.999{\times}10^{-1}$ & 0.91 & 1.06 & 1.77 \\
2 & 2.500 & 0.750 & $-0.1808$ & 0.0170 & 0.0081 & $-0.0780$ & 0.9975 & 0.7896 & 1.0032 & $3.964{\times}10^{-1}$ & 0.47 & 1.06 & 1.77 \\
2 & 2.500 & 0.820 & $-0.1805$ & 0.0170 & 0.0076 & $-0.0828$ & 0.9978 & 0.7859 & 1.0028 & $6.634{\times}10^{-1}$ & 0.61 & 1.06 & 1.77 \\
2 & 2.500 & 0.913 & $-0.1875$ & 0.0170 & 0.0062 & $-0.0933$ & 0.9983 & 0.7715 & 1.0016 & $1.782{\times}10^{+0}$ & 0.64 & 1.06 & 1.77 \\
2 & 3.250 & 0.426 & $-0.1925$ & 0.0186 & 0.0119 & $-0.0654$ & 0.9962 & 0.7936 & 1.0077 & $2.870{\times}10^{-2}$ & 1.23 & 1.06 & 1.77 \\
2 & 3.250 & 0.609 & $-0.1870$ & 0.0186 & 0.0105 & $-0.0710$ & 0.9973 & 0.7932 & 1.0049 & $6.817{\times}10^{-2}$ & 0.88 & 1.06 & 1.77 \\
2 & 3.250 & 0.719 & $-0.1854$ & 0.0186 & 0.0095 & $-0.0767$ & 0.9977 & 0.7896 & 1.0042 & $1.261{\times}10^{-1}$ & 0.86 & 1.06 & 1.77 \\
2 & 3.250 & 0.865 & $-0.1926$ & 0.0186 & 0.0074 & $-0.0900$ & 0.9982 & 0.7723 & 1.0021 & $3.906{\times}10^{-1}$ & 0.48 & 1.06 & 1.77 \\
2 & 4.000 & 0.437 & $-0.1923$ & 0.0199 & 0.0140 & $-0.0636$ & 0.9968 & 0.7984 & 1.0084 & $1.308{\times}10^{-2}$ & 1.43 & 1.06 & 1.77 \\
2 & 4.000 & 0.593 & $-0.1901$ & 0.0199 & 0.0120 & $-0.0703$ & 0.9975 & 0.7938 & 1.0058 & $2.786{\times}10^{-2}$ & 1.25 & 1.06 & 1.77 \\
2 & 4.000 & 0.694 & $-0.1888$ & 0.0199 & 0.0109 & $-0.0754$ & 0.9979 & 0.7902 & 1.0045 & $4.951{\times}10^{-2}$ & 1.25 & 1.06 & 1.77 \\
2 & 4.000 & 0.805 & $-0.1981$ & 0.0199 & 0.0085 & $-0.0866$ & 0.9981 & 0.7725 & 1.0030 & $1.026{\times}10^{-1}$ & 0.89 & 1.06 & 1.77 \\
2 & 4.000 & 0.946 & $-0.2431$ & 0.0199 & 0.0038 & $-0.1111$ & 0.9983 & 0.7174 & 1.0009 & $6.186{\times}10^{-1}$ & 0.76 & 1.06 & 1.77 \\
2 & 5.000 & 0.389 & $-0.1950$ & 0.0214 & 0.0174 & $-0.0601$ & 0.9970 & 0.8031 & 1.0094 & $4.245{\times}10^{-3}$ & 2.06 & 1.06 & 1.77 \\
2 & 5.000 & 0.538 & $-0.1942$ & 0.0214 & 0.0140 & $-0.0676$ & 0.9977 & 0.7956 & 1.0069 & $8.521{\times}10^{-3}$ & 1.46 & 1.06 & 1.77 \\
2 & 5.000 & 0.704 & $-0.2029$ & 0.0214 & 0.0105 & $-0.0807$ & 0.9980 & 0.7759 & 1.0043 & $2.137{\times}10^{-2}$ & 1.05 & 1.06 & 1.77 \\
2 & 5.000 & 0.919 & $-0.2472$ & 0.0214 & 0.0044 & $-0.1124$ & 0.9982 & 0.7149 & 1.0010 & $1.578{\times}10^{-1}$ & 1.04 & 1.06 & 1.77 \\
2 & 6.000 & 0.886 & $-0.2516$ & 0.0226 & 0.0049 & $-0.1120$ & 0.9981 & 0.7132 & 1.0010 & $4.754{\times}10^{-2}$ & 1.24 & 1.06 & 1.77 \\
2 & 7.000 & 0.847 & $-0.2568$ & 0.0237 & 0.0053 & $-0.1113$ & 0.9980 & 0.7110 & 1.0023 & $1.711{\times}10^{-2}$ & 2.26 & 1.06 & 1.77 \\
3 & 1.750 & 0.250 & $-0.1880$ & 0.0149 & 0.0086 & $-0.0457$ & 0.9958 & 0.8071 & 1.0083 & $1.527{\times}10^{-1}$ & 0.21 & 1.12 & 1.77 \\
3 & 2.500 & 0.227 & $-0.1885$ & 0.0170 & 0.0127 & $-0.0407$ & 0.9966 & 0.8163 & 1.0101 & $3.581{\times}10^{-2}$ & 0.28 & 1.12 & 1.77 \\
3 & 3.250 & 0.206 & $-0.1957$ & 0.0186 & 0.0156 & $-0.0387$ & 0.9968 & 0.8159 & 1.0115 & $1.108{\times}10^{-2}$ & 0.67 & 1.12 & 1.77 \\
3 & 4.000 & 0.190 & $-0.2018$ & 0.0199 & 0.0184 & $-0.0368$ & 0.9969 & 0.8159 & 1.0123 & $4.142{\times}10^{-3}$ & 0.81 & 1.12 & 1.77 \\
3 & 5.000 & 0.171 & $-0.1990$ & 0.0214 & 0.0250 & $-0.0314$ & 0.9973 & 0.8297 & 1.0141 & $1.358{\times}10^{-3}$ & 0.93 & 1.12 & 1.77 \\
3 & 6.000 & 0.156 & $-0.2029$ & 0.0226 & 0.0297 & $-0.0286$ & 0.9974 & 0.8338 & 1.0150 & $5.241{\times}10^{-4}$ & 1.27 & 1.12 & 1.77 \\
3 & 7.000 & 0.143 & $-0.2048$ & 0.0237 & 0.0349 & $-0.0255$ & 0.9977 & 0.8403 & 1.0165 & $2.285{\times}10^{-4}$ & 2.26 & 1.12 & 1.77 \\
3 & 8.830 & 0.125 & $-0.2208$ & 0.0254 & 0.0351 & $-0.0259$ & 0.9974 & 0.8280 & 1.0217 & $6.153{\times}10^{-5}$ & 3.89 & 1.12 & 1.77 \\
\end{tabular}
\end{ruledtabular}
\end{table*}

The ninth column of Table~\ref{Tab1} lists the new RC factors $C_{\text{rad}}^{\text{new}}$, calculated in accordance with Eq.~(\ref{Eq12}). The next column shows the ratios $C_{\text{rad}}^{\text{old}} / C_{\text{rad}}^{\text{new}}$, ranging from 1.0009 to 1.0217. These are the overall correction factors by which we multiply the differential cross sections reported in Refs.~\cite{PRD.49.5671, PRD.50.5491}. The values of $C_{\text{rad}}^{\text{old}} / C_{\text{rad}}^{\text{new}}$ increase with decreasing~$\varepsilon$ and are especially large for Set~3 where $\theta = 90^{\circ}$. The new, corrected values of $d \sigma_0 / d \Omega$ are also given in Table~\ref{Tab1}. Finally, its last three columns list the statistical, point-to-point systematic, and overall normalization uncertainties of the cross sections.

The overall normalization uncertainties are 1.9\% for Set~1 and 1.77\% for the other two sets. These are due to the absolute normalization of the cross sections and are completely correlated for the measurements within the same set. All the uncertainties reported here are taken directly from the original analyses. The only difference is that the authors of Ref.~\cite{PRD.50.5491} assigned an additional point-to-point systematic uncertainty of 0.7\% to the cross sections of Set~3 because these were normalized to the overlapping points from Set~2. The normalization factor found was $0.958 \pm 0.007$. In our extraction of the proton form factors, we normalize the three data sets simultaneously.

In the original analyses, the normalization uncertainty due to RCs was estimated as~1\%. The systematic uncertainty of 0.5\% was additionally introduced in Ref.~\cite{PRD.50.5491} to account for inaccuracies in the RCs. Therefore, some of the corrections we made to $d \sigma_0 / d \Omega$ are beyond the RC uncertainties claimed in Refs.~\cite{PRD.49.5671, PRD.50.5491}.

Let us now describe our extraction of the proton form factors, which is different from the procedures used in the original analyses. First of all, we assign to each data set a single normalization factor which multiplies all the cross sections within the set. Then, the usual least-squares technique is used to fit the cross-section data. The function to minimize is
\begin{gather}
\chi^2 = \sum\limits_{i=1}^{3} \sum\limits_{j=1}^{N_i} \frac{\left[n_i \sigma_{ij} - \varepsilon_{ij} G_E^2 (\tau_{ij}) - \tau_{ij} G_M^2 (\tau_{ij})\right]^2}{(\Delta \sigma_{ij})^2} \nonumber \\
{} + \sum\limits_{i=1}^{3} \frac{(n_i - 1)^2}{(\Delta n_i)^2},
\end{gather}
where $n_i$~are the three unknown normalization factors, $\Delta n_i$ are the corresponding normalization uncertainties, $\sigma_{ij}$~are the fitted reduced cross sections, and $\Delta \sigma_{ij}$~are the combined statistical and systematic uncertainties of~$\sigma_{ij}$. Here, $i = 1$, 2, and 3 enumerates the data sets and $j$~enumerates the kinematics within each set.

\begin{table}
\begin{ruledtabular}
\caption{\label{Tab2}Values of the best-fit parameters.}
\begin{tabular}{lrrr}
& \multicolumn{1}{c}{$i = 1$} & \multicolumn{1}{c}{$i = 2$} & \multicolumn{1}{c}{$i = 3$} \\
\hline $n_i$ & $1.012 \pm 0.011$ & $1.014 \pm 0.011$ & $0.975 \pm 0.011$ \\
$a_i$ & $0.197 \pm 0.211$ & $0.703 \pm 0.426$ & $-0.454 \pm 0.209$ \\
$b_i$ & $-0.444 \pm 0.043$ & $0.397 \pm 0.045$ & $-0.081 \pm 0.013$ \\
\end{tabular}
\end{ruledtabular}
\end{table}

The minimization can be done in an especially simple and elegant way if we choose the following parametrization for $G_E^2$ and~$G_M^2$:
\begin{gather}
G_E^2 (\tau) = \left(1 - a_1 \tau - a_2 \tau^2 - a_3 \tau^3\right) G_D^2 (\tau), \label{Eq44} \\
G_M^2 (\tau) = \left(1 - b_1 \tau - b_2 \tau^2 - b_3 \tau^3\right) \mu^2 G_D^2 (\tau), \label{Eq45}
\end{gather}
where $a_i$ and $b_i$ are six unknown parameters and $G_D$ is the dipole form factor. Although the functions (\ref{Eq44})--(\ref{Eq45}) do not satisfy the asymptotic behavior $G_{E,M} \propto \tau^{-2}$ expected from dimensional scaling laws in perturbative QCD~\cite{PRD.11.1309}, they have the advantage of being linear in the unknowns $a_i$ and~$b_i$. Therefore, if we calculate and then set to zero the partial derivatives $\partial \chi^2 / \partial n_i$, $\partial \chi^2 / \partial a_i$, and $\partial \chi^2 / \partial b_i$, we obtain nine linear equations to determine the nine free parameters.

\begin{figure}
\includegraphics[width=\columnwidth]{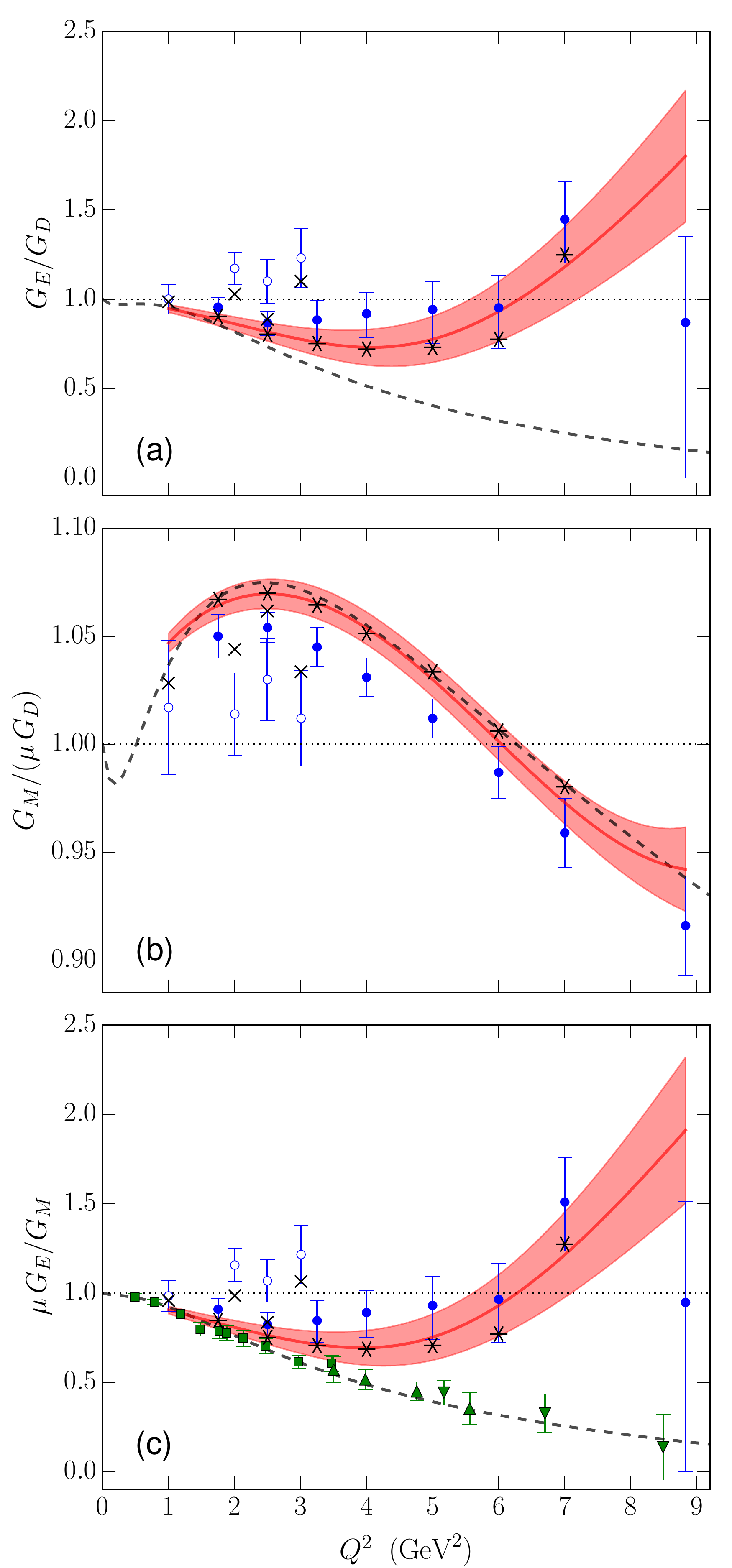}
\caption{\label{Fig2}Results of the reanalysis in comparison with the original data for (a) $G_E / G_D$, (b) $G_M / (\mu \, G_D)$, and (c) $\mu \, G_E / G_M$ as functions of~$Q^2$. The red solid lines represent the best fit to the cross-section data obtained using the parametrization (\ref{Eq44})--(\ref{Eq45}), while the shaded areas are the 68\% confidence bands. The blue open and solid circles are the original data of Refs.~\cite{PRD.49.5671} and~\cite{PRD.50.5491}, respectively. The black crosses (asterisks) illustrate how the individual data points reported in Ref.~\cite{PRD.49.5671} (Ref.~\cite{PRD.50.5491}) change after reanalysis. The gray dashed lines correspond to the Kelly parametrization~\cite{PRC.70.068202}. Results of the polarization transfer measurements \cite{PRC.71.055202} (green squares), \cite{PRC.85.045203} (green triangles), and~\cite{PRL.104.242301} (green inverted triangles) are also shown in panel~(c).}
\end{figure}

These linear equations can be written in a matrix form as $A \mathbf{x} = \mathbf{b}$, where $A$ is a real symmetric $9 \times 9$ matrix and $\mathbf{x}, \mathbf{b} \in \mathbb{R}^9$. The vector~$\mathbf{x}$ of the best-fit parameters is calculated as $\mathbf{x} = A^{-1} \mathbf{b}$. Note that the matrix $A^{-1}$, inverse to~$A$, is conveniently the covariance matrix providing information about the uncertainties of and the correlations between the fitted parameters.

The whole analysis procedure was repeated iteratively to take into account the dependence of the bremsstrahlung corrections~(\ref{Eq28}) and~(\ref{Eq34}) on the form factor parametrization used to calculate $d \sigma_0 / d \Omega$. We started from the dipole form factors ($a_i = 0$, $b_i = 0$) and then obtained subsequent values of~$a_i$ and~$b_i$. The procedure converged after only a few iterations.

The resulting best-fit parameters and their uncertainties are given in Table~\ref{Tab2}. The corresponding chi-square value is $\chi^2 = 26.1$ for $N_1 + N_2 + N_3 - 9 = 39$ degrees of freedom. Figure~\ref{Fig2} compares our results with the original ones and with the Kelly fit~\cite{PRC.70.068202}, which accounts for some polarization transfer measurements. The data points reported in Refs.~\cite{PRD.49.5671} and~\cite{PRD.50.5491} are shown by the blue open and solid circles, respectively. The error bars correspond to the combined statistical and systematic uncertainties. The black crosses and asterisks illustrate how the original results of Refs.~\cite{PRD.49.5671} and~\cite{PRD.50.5491} change after reanalysis. These were obtained with the standard Rosenbluth separation technique using the corrected cross sections multiplied by the normalization factors listed in Table~\ref{Tab2}. Note that for $Q^2 = 8.83~\text{GeV}^2$ there is only one value of $d \sigma_0 / d \Omega$ measured and the Rosenbluth method cannot be applied without using third-party data, as was done in Ref.~\cite{PRD.50.5491}.

The red solid lines in Fig.~\ref{Fig2} represent the best fit to the corrected cross sections. The shaded areas are the corresponding 68\% confidence bands calculated using the uncertainty propagation method and taking into account correlations between the fitted parameters. By choosing a specific form factor parametrization we introduced a model bias that is another source of uncertainty. We expect that this effect is not significant because the polynomial model~(\ref{Eq44})--(\ref{Eq45}) is flexible and the best fit is in good agreement with the Rosenbluth extraction. However, our results for $Q^2 > 7~\text{GeV}^2$, where only one cross section is available, should be interpreted with caution.

As can be seen from Fig.~\ref{Fig2}, at $Q^2 \leqslant 7~\text{GeV}^2$ our analysis gives for $G_E$ and $G_E / G_M$ slightly lower values than those extracted previously. At the same time, the new values of~$G_M$ are consistently higher and thus closer to the Kelly parametrization. The original data from Ref.~\cite{PRD.49.5671} is in poor agreement with the more precise measurement~\cite{PRD.50.5491} and appear to be less reliable.

\section{Conclusion}
\label{Sec5}

We have reanalyzed the data from the SLAC measurements~\cite{PRD.49.5671, Walker_thesis, PRD.50.5491, Clogher_thesis} in light of the discrepancy between the Rosenbluth and polarization transfer methods. The corresponding RCs were revisited taking into account recent theoretical developments in this field. We followed the RC procedure proposed by Walker \textit{et~al.}~\cite{PRD.49.5671}, but corrected misprints and inaccuracies that could possibly affect the results of the original analyses. We calculated the standard internal RCs in accordance with the Maximon--Tjon prescription~\cite{PRC.62.054320}, which is an improvement over the previously used Mo--Tsai formalism~\cite{RMP.41.205}. The revised formulas and their {\scshape python} implementation~\cite{GitHub} may be useful for future single-arm measurements of unpolarized elastic electron-proton scattering.

The new values of $d \sigma_0 / d \Omega$ obtained after reapplying RCs are listed in Table~\ref{Tab1}. They are higher than the original values by an amount of 0.09\% to 2.17\%. Using the corrected cross sections, we determined the proton electric and magnetic form factors in the $Q^2$ range from 1~to $8.83~\text{GeV}^2$. The parametrization we chose for~$G_E^2$ and~$G_M^2$ is given by Eqs.~(\ref{Eq44})--(\ref{Eq45}), and the best-fit parameters found are shown in Table~\ref{Tab2}.

Our extraction of the proton form factors differs from the standard Rosenbluth separation technique. The procedure we used does not require measuring two or more cross sections at the same $Q^2$~value. The only apparent disadvantage of this approach is the need to assume a specific form factor parametrization.

Finally, the detailed reanalysis of the Rosenbluth measurements~\cite{PRD.49.5671, PRD.50.5491} brings their combined results into better agreement with the polarization transfer data. Our results confirm a significant experimental discrepancy from polarization measurements for $Q^2 \agt 3~\text{GeV}^2$, but not at lower $Q^2$.

\begin{acknowledgments}
The authors are grateful to Dr. E.~Tomasi-Gustafsson for discussions and her stimulating interest in the subject. This work was supported by the Ministry of Education and Science of the Russian Federation.
\end{acknowledgments}

\end{document}